\documentclass[jkps,final,twocolumn,fleqn,showpacs,showkeys]{revtex4-1}
\usepackage[pdftex]{graphicx}
\usepackage{amssymb}
\usepackage{amsmath}
\usepackage{bm}
\usepackage{multirow}

\begin{document}
\preprint{J. Korean Phys. Soc., Vol. 61, No. 10, November 2012, pp. 1609-1614}
\title[Structural changes in SnO$_2$ and TiO$_2$ thin films by SHI
irradiation]{On the Optical Properties of Ag$^{+15}$ Ion-beam-irradiated TiO$_2$
and SnO$_2$ Thin Films}
\author{Hardeep \surname{Thakur}} \author{K. K. \surname{Sharma}} \author{Ravi \surname{Kumar}}
\affiliation{Center for Material Science and Engineering, National Institute of Technology, Hamirpur 177-005, India}
\author{Pardeep \surname{Thakur}}
\altaffiliation{Diamond Light Source Ltd., Didcot, Oxfordshire, OX11 0DE, UK}
\affiliation{European Synchrotron Radiation Facility, BP 220, F-38043 Grenoble Cedex, France}
\author{Yogesh \surname{Kumar}} \author{Abhinav Pratap \surname{Singh}}
\affiliation{Material Science Division, Inter University Accelerator Center, New Delhi 110-067, India}
\author{Sanjeev \surname{Gautam}}
\email{sgautam71@kist.re.kr} \thanks{Phone/Fax: +82-54-279-1192/1599}
\author{Keun Hwa \surname{Chae}}
\affiliation{Advanced Analysis Center, Korea Institute of Science and Technology (KIST), Seoul 136-791, Republic of Korea}
\received{December 2 2011}

\begin{abstract}
The effects of 200-MeV Ag$^{+15}$ ion irradiation on the optical properties of TiO$_2$ and SnO$_2$ thin films prepared by using the RF magnetron sputtering technique were investigated. These films were characterized by using UV-vis spectroscopy, and with increasing  irradiation fluence, the transmittance for the TiO$_2$ films was observed to increase systematically while that for SnO$_2$ was observed to decrease. Absorption spectra of the irradiated samples showed minor changes in the indirect bandgap from $3.44$ to $3.59$ eV with increasing irradiation fluence for TiO$_2$ while significant
changes in the direct bandgap from $3.92$ to $3.6$ eV were observed for SnO$_2$. The observed modifications in the optical properties of both the TiO$_2$ and the SnO$_2$ systems with irradiation can be attributed to controlled structural disorder/defects in the system.
\end{abstract}

\keywords{SHI irradiation, Optical properties, UV-vis, Surface modifications\\ DOI:10.3938/jkps.61.1609}
\maketitle

\section{Introduction}
\label{intro} Due to increasing interest in electronics and optoelectronics, among the wide-band-gap semiconductors, TiO$_2$ and SnO$_2$ are being considered as the most promising materials in view of their unique properties and various future technological applications. These applications boast their moderate price, high-volume, nontoxicity and chemical stability. In addition, these materials offer the possibility of integrating their magnetic and electronic properties in spintronic devices by using both the spin and the charge of electrons \cite{r1,r2}.

TiO$_2$ is a very interesting and versatile material with a wide range of applications, including use in microelectronics due to its high dielectric constant and in optical coatings due to its high refractive index \cite%
{r3,r4,r5,r6,r7,r8}. It also has excellent optical transmittance in the
visible and the near-infrared regions. TiO$_2$ exists in three crystalline
polymorphs: rutile, anatase, and brookite with band gap values of $3.03$,
$3.19$, and $3.11$ eV, respectively \cite{r9}. Among the different TiO$_2$
polymorphs, anatase (tetragonal, $D^{19}_{4h}$) is a metastable phase that
contains four shared edges per octahedron (the highest condensation of TiO$_6
$ octahedra) and is known to be useful for photocatalysis with response to
ultraviolet photons. The rutile (tetragonal, $D^{14}_{4h}$) is
the thermodynamically most stable phase at all temperatures and is formed by
sharing two edges per octahedron (the lowest condensation of TiO$_6$
octahedra) with the largest index of refraction. The brookite (orthorhombic, $%
D^{15}_{4h}$) is the most distorted phase and shares three edges per
octahedron. The properties of TiO$_2$  significantly depend on the
crystalline phases, i.e., anatase, rutile, or brookite, and on the morphology of
the material \cite{r10}.

On the other hand, tin dioxide (SnO$_2$) has been investigated in the view of
potential technological applications in catalysis, gas sensor technology,
etc. \cite{r11,r12,r13} because of the high carrier density, optical
transparency, wideband gap ($\sim 3.6$ eV), and remarkable chemical and
thermal stabilities. SnO$_2$ exists in the most important form of
the crystalline phase, known as cassiterite, with a rutile (tetragonal, $%
D^{14}_{4h}$) structure. Another form of SnO$_2$ with an orthorhombic
structure is known to be stable only at high pressures and temperatures.

Many deposition techniques (pulsed laser deposition, sol-gel deposition, etc.) have been employed to synthesize TiO$_2$ and SnO$_2$ thin films, although magnetron sputtering remains the preferred method due to its better coating uniformity, process versatility, large-area coating capability and more freedom in selecting the deposition conditions \cite{r14}. Any mechanism that affects the lattice structure of the TiO$_2$ and the SnO$_2$ systems also influences their electronic structures and optical properties. Swift heavy ion (SHI) irradiation is one of the mechanisms that have been used to tailor the material properties by modifying its electronic structure \cite{r14,r15,r16}. Radiation-induced defects produced in the material are well documented to depend entirely upon the energy loss processes, namely, nuclear energy loss (elastic process), and electronic energy loss
(inelastic process), involved during the passage of ions through the target material. In the high-energy region, due to dense electronic excitations, SHIs induce point/cluster/columnar defects and structural disorder, depending upon the extent of the electronic energy loss mechanism in the system. Previous studies on the TiO$_2$ and the SnO$_2$ systems only focused on the SHI-induced modifications in the electronic structure, orbital anisotropy and magnetic properties \cite{r15,r16}. Thus, other properties, such as the optical properties, need further study. This experimental study was conducted to observe the changes in the optical transmittance, absorption and band gaps of TiO$_2$ and SnO$_2$ thin films irradiated with a 200-MeV Ag$^{+15}$ ion beam at various irradiation fluences ranging from $1\times10^{11}$ to $5\times10^{12}$ ions/cm$^2$ which is different from what is largely reported.

\section{Experiment and Discussion}
Pure titanium (II) oxide (TiO) and SnO$_{2}$ compounds (purity 99.9\%) were used as the starting materials for the depositions of the thin films. TiO$_{2}$ and SnO$_{2}$ films, approximately $\sim 100$-nm thick, were deposited on cleaned sapphire single-crystal substrates by using a RF magnetron sputtering technique. The pure oxide materials were ground into a fine powder in an agate mortar; then, mixtures were pressed in the form of circular targets of $50$ mm in diameter by applying a pressure of $5-6$ tons in a
hydraulic press. The targets were sintered at $1000^{\circ }$C for $\sim 12$ h. Prior to filling the chamber with sputtering gas, it was evacuated to a base pressure of $\sim 1.1\times 10^{-5}$ Torr by using a turbo-molecular pump. The deposition was carried out in a partial pressure of $10$ mTorr of an oxygen and Ar gas mixture (1:1), keeping the substrate temperature at $550^{\circ}$C and RF power at $100$ W. After deposition, the films were annealed in-situ at $550^{\circ }$C in oxygen for $1$ h. The deposited thin films were irradiated with 200-MeV Ag$^{+15}$ ions at fluences of $1\times 10^{11}$, $%
1\times 10^{12}$, and $5\times 10^{12}$ ions/cm$^{2}$ at room temperature
(RT) by using the 15UD tandem accelerator at the Inter-University Accelerator Center, New Delhi, India.
\begin{figure}[tbh] \centering
\includegraphics[width=6.5cm]{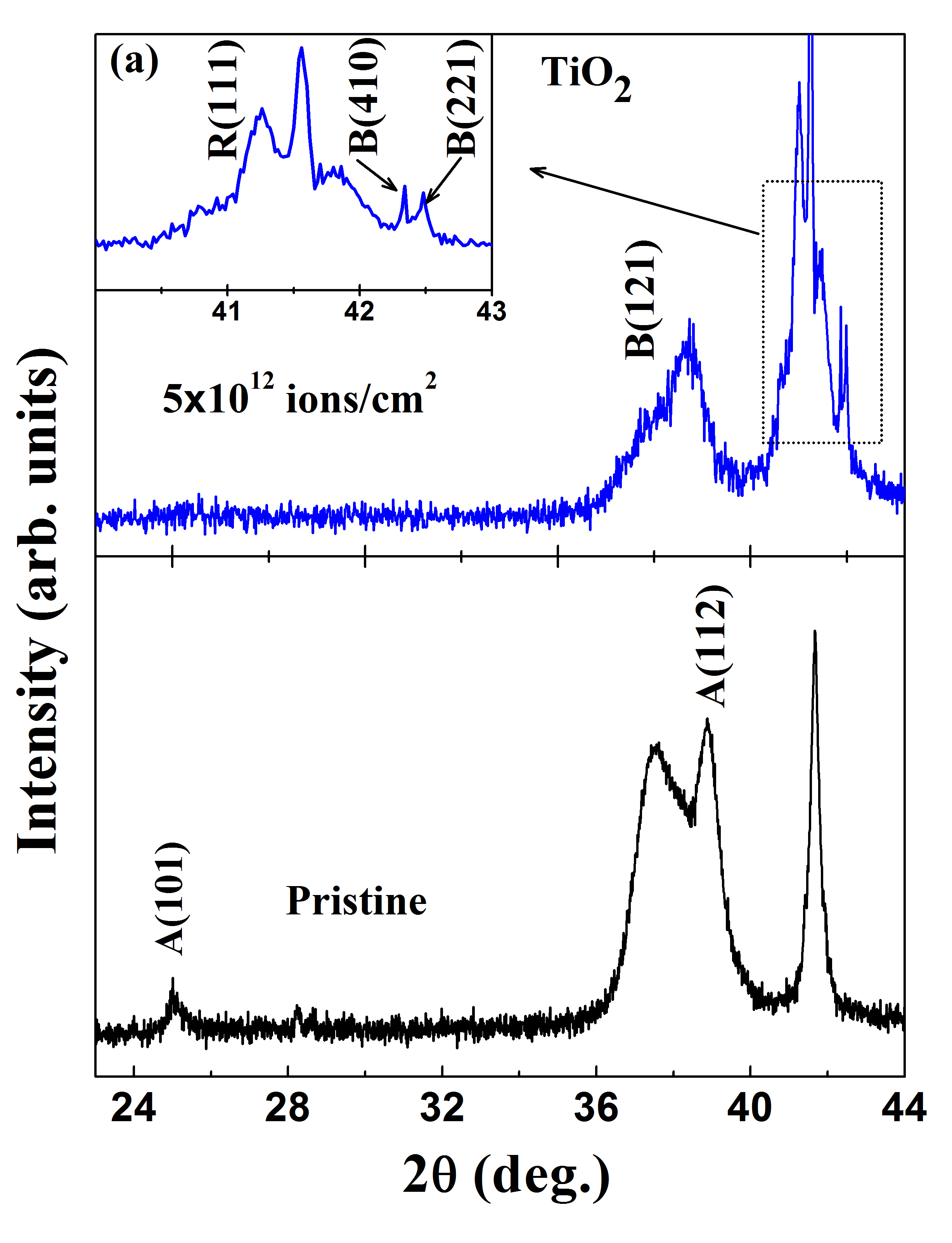}\\  %
\includegraphics[width=6.5cm]{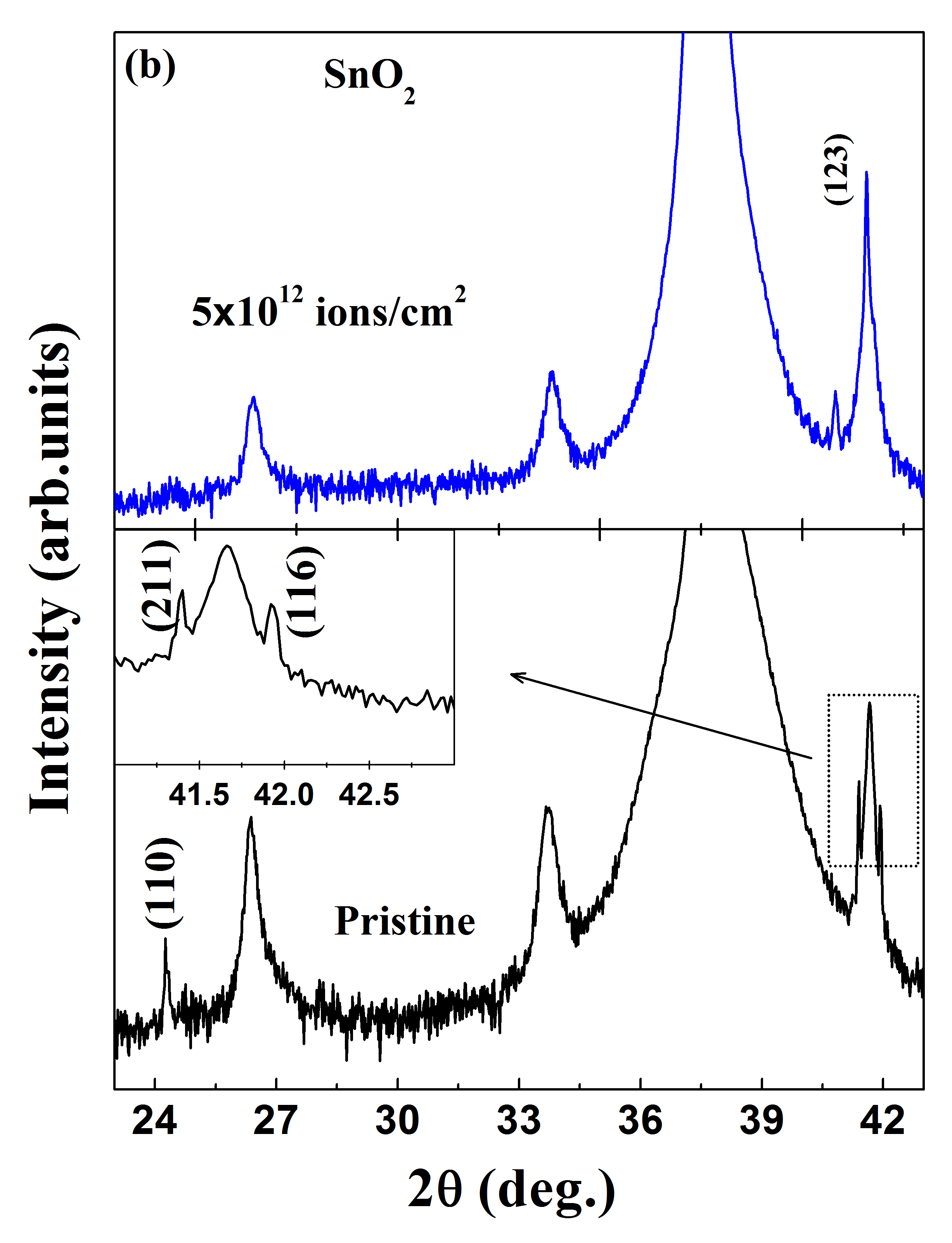}\\
\caption{(Color online) HRXRD pattern of (a) pristine and irradiated (SHI: $5\times 10^{12}$
ions/cm$^{2}$) TiO$_{2}$ thin films and (b) pristine and irradiated SnO$_{2}$
thin films. The inset in Fig. \protect\ref{fig1}(a) shows an extended view
of the mixed brookite and rutile phases of the irradiated film and that in (b)
shows an extended view of the orthorhombic phase of SnO$_{2}$.} \label{fig1}
\end{figure}

The structural analysis of the pristine and the irradiated TiO$_{2}$ and SnO$_{2}
$ films was carried out using high-resolution X-ray diffraction (HRXRD) with $\lambda =1.5425 \AA$ at the bending magnet 1D XRS KIST-PAL beamline of the Pohang Accelerator Laboratory (PAL), Korea. The HRXRD profiles are shown in Fig.~ \ref{fig1}(a) for the pristine and the SHI irradiation TiO$_{2}$ thin films. As shown in Fig. \ref{fig1}(a), the pristine sample has a tetragonal anatase structure (JCPDS, Card No.84-1286) while on irradiation at the highest SHI fluence of $5\times 10^{12}$ ions/cm$^{2}$, the inset in
Fig. \ref{fig1}(a) clearly shows mixed peaks of the brookite (JCPDS, Card No.76-1937) and the rutile phases of TiO$_{2}$. The appearance of broader brookite peaks clearly indicates that SHI irradiation has induced structural disorder and/or strain in the films \cite{r16}.
Figure \ref{fig1}(b) shows HRXRD pattern of the pristine and the SHI-irradiated SnO$_{2}$
thin films collected at RT. For the pristine sample depicted in Fig. \ref{fig1}(b), it is evident that the characteristic peaks at $2\theta = 24.20^{\circ}, 41.40^{\circ}$, and $41.90^{\circ}$ correspond to reflections from the (110), (211), and (116) planes of the orthorhombic phase (JCPDS, Card No.78-1063) of SnO$_{2}$, respectively. The inset in Fig. \ref{fig1}(b) shows the extended view of the diffraction peaks of the pristine sample. At
the highest SHI fluence of $5\times 10^{12}$ ions/cm$^{2}$, the irradiation causes partial amorphization and/or strain in the SnO$_{2}$ system. The detailed structural verification of the SnO$_{2}$ orthorhombic structure and the effect of SHI irradiation on its structure is discussed elsewhere \cite{r18}.
\begin{figure}[tbh!]\centering
\includegraphics[width=6.5cm]{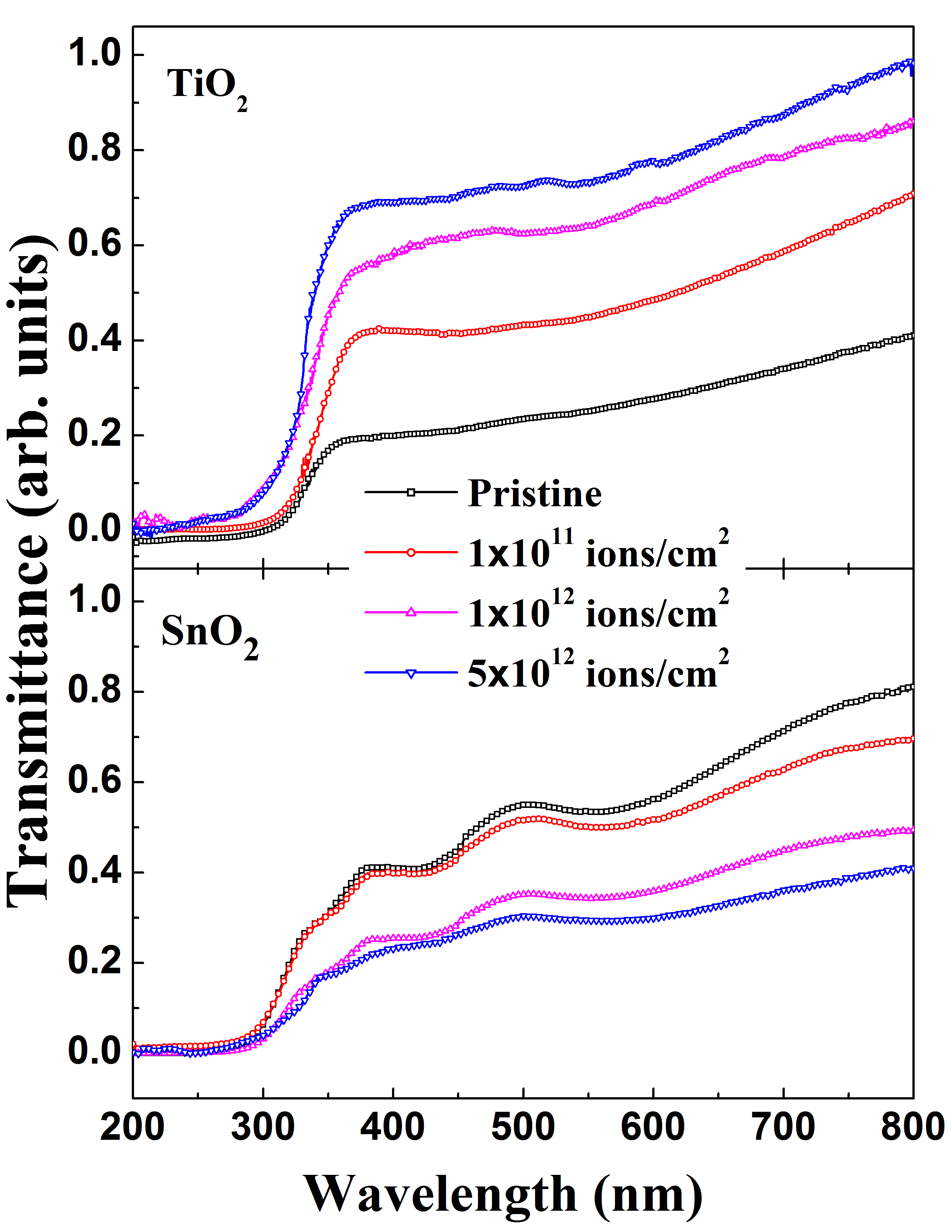}\\[-0.5cm]
\caption{(Color online)Transmittance spectra for the pristine and the irradiated
(SHI: $1\times10^{11}-5\times10^{12}$ ions/cm$^2$) TiO$_2$ (upper panel) and
SnO$_2$ (bottom panel) thin films collected at RT.}\label{fig2}
\end{figure}

The optical transmittances of the pristine and the irradiated thin films of both
oxides were measured by using a Hitachi-U3300 spectrophotometer to collect the transmittance spectra as a function
of wavelength in the range of $200-800$ nm at RT with a resolution of $\lambda = 0.5$ nm.
Figure \ref{fig2} presents optical transmittance spectra for the pristine
and the SHI-irradiated TiO$_2$ (upper panel) and SnO$_2$ (bottom panel) thin
films collected at RT. It is clear from Fig. \ref{fig2} that the pristine TiO$%
_2$ film exhibits a transmittance value of about 25\% in the visible region
at $\sim 470$ nm. With increasing irradiation (SHI: $1\times10^{11}-
5\times10^{12}$ ions/cm$^2$) fluence, the transmittance increases
considerably, which at highest fluence (SHI: $5\times10^{12}$ ions/cm$^2$)
acquires a value of $\sim$ 70\%. The increase in the transmittance with increasing SHI
fluence signifies that the transparency of the SHI-irradiated TiO$_2$ films is
superior to that of the pristine film. On the other hand, the pristine SnO$_2
$ film exhibits a systematic decrease in the transmittance with increasing
irradiation (SHI: $1\times10^{11}-5\times10^ {12}$ ions/cm$^2$) fluence. In
the visible region, the transmittance for the pristine SnO$_2$ film was
$\sim$ 50\% and decreased to $\sim$35\% for the highest SHI fluence.
Moreover, at $\sim 350$ nm, the transmittance decreases quickly for all samples for both the oxides materials and approaches zero at $\sim 300$ nm. This fast decrease in the transmittance is due to strong absorption of
light in this region caused by the excitation and the migration of electrons
from the valence band to the conduction band. Typical oscillations in the
transmittance spectra, particularly for SnO$_2$, may be due to interference of
light transmitted through the thin film and the substrate \cite{r19}.

Previous studies \cite {r16,r17,r18} revealed that the irradiated TiO$_2$
thin films for the highest SHI fluence exhibited a mixed (dominating brookite
+ rutile) phase of TiO$_2$ while the SnO$_2$ pristine film was composed of a
pure orthorhombic phase of SnO$_2$, and SHI-induced controlled structural
disorder (distortion in the SnO$_6$ octahedra) and/or strain in the films.
The dominating structure of the brookite phase in the SHI-irradiated TiO$_2$
films and the orthorhombic distortions of the SnO$_2$ lattice could have important implications for the
electronic structure and possible optical properties. Therefore, the observed changes in the transmittance of the
pristine TiO$_2$ and SnO$_2$ films with SHI fluence can be attributed to
TiO$_6$ and SnO$_6$ octahedral distortions, respectively.

\begin{figure}[tbh!]\centering
\includegraphics[width=6.5cm]{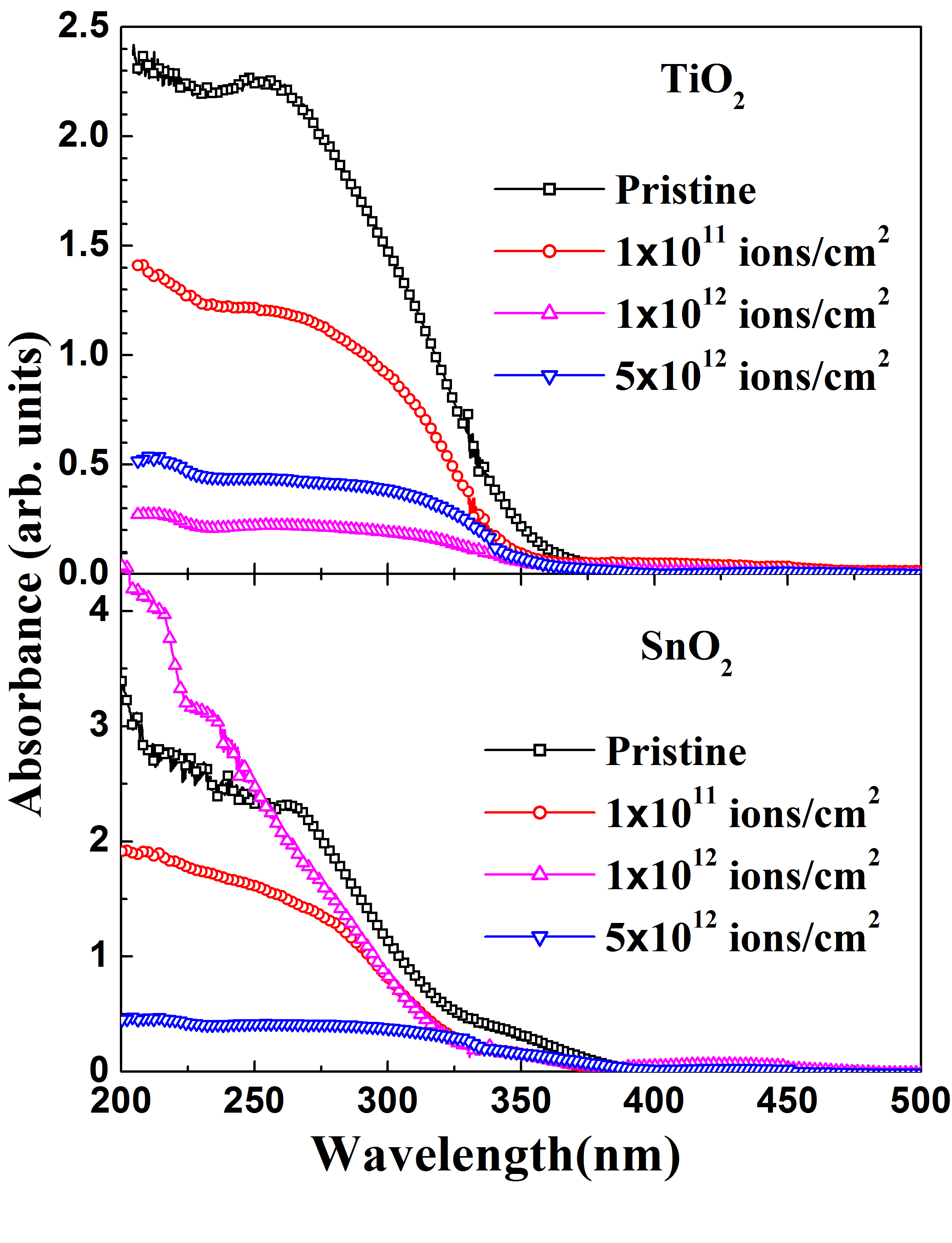}\\[-0.5cm]
\caption{(Color online) Optical absorption spectra for the pristine and the irradiated
(SHI: $1\times10^{11}-5\times10^{12}$ ions/cm$^2$) TiO$_2$ (upper panel) and
SnO$_2$ (bottom panel) thin films collected at RT.}\label{fig3}
\end{figure}

The absorption spectra for the pristine and the irradiated TiO$_2$ and SnO$_2$
films were measured by measuring the absorbance as a function of wavelength
at RT. The absorbance is given by
\begin{equation}
A = \log \left( \frac{I}{I_0} \right)  \label{eq1}
\end{equation}
where $I_0$ is the intensity of the incident radiation and $I$ is the transmitted intensity.
Figure \ref{fig3} depicts the absorption spectra for the pristine and the
SHI-irradiated TiO$_2$ (upper panel) and SnO$_2$ (bottom panel) thin films. The
optical absorption spectra of TiO$_2$ showed a clear absorption edge at $\sim
 352$ nm for the pristine TiO$_2$ sample. With increasing irradiation
(SHI: $1\times10^{11}-5\times10^{12}$ ions/cm$^2$) fluence, the absorption edge is
slightly shifted to the smaller wavelength side, and at the highest
irradiation (SHI: $5\times10^{12}$ ions/cm$^2$) fluence, it acquires a value of
$\sim 345$ nm. The observed values of the absorption edge deviate from the
values reported in the literature \cite{r9}. For the SnO$_2$ system, the
optical absorption edge of the pristine film was found to be at $\sim 335$
 nm, which is shifted to $\sim 325$ nm at the highest SHI fluence. In this
study, an anomalous trend was observed for the optical absorption edges and
the suppression in the maximum absorption with increasing SHI fluence for
both the oxide materials. The absorption spectra of both systems reveal
that the films grown under the same parametric conditions have low
absorbance in the visible/near-infrared region while the absorbance is high in
the ultraviolet region. Since SHI induces a controlled structural disorder
(TiO$_6$ and SnO$_6$ octahedral distortions) in both the systems,
modifications in the absorption spectra can be correlated to changes in the
electronic structure as a result of a lowering in the orbital symmetry (i.e.,
its $s$-, $p$-, and $d$-like character) via strong hybridization effects
after irradiation.

The absorption coefficient as a function of photon energy for
allowed indirect transitions \cite{r17} is given by
\begin{equation}
\alpha = B_i (h\nu -E_g \pm E_p)^2  \label{eq2}
\end{equation}
and for allowed direct transitions by
\begin{equation}
\alpha = B_d (h\nu -E_g)^{1/2},  \label{eq3}
\end{equation}
where $\alpha$ is the absorption coefficient, $B_i$ and $B_d$ are constants for indirect and direct transitions, $h$ is Planck's constant, $\nu$ is the frequency, $E_p$ is the photon energy involved in the indirect transition, and $E_g$ is the band gap energy. The absorption coefficient $\alpha$ is obtained from Beer's law:
\begin{equation}
I=I_0 \exp (-\alpha t)  \label{eq4}
\end{equation}
In Eq. \ref{eq4}, $t$ is the thickness of the measured sample.

The relationship between the absorbance $A$, the absorption coefficient $\alpha$,
and the thickness of the film $t$ is given by
\begin{equation}
\alpha =2.303 \left( \frac{A}{t}. \right)  \label{eq5}
\end{equation}
A plot of $\alpha^{1/2}$ versus energy was used to obtain the value of
the indirect band gap, and a plot of $\alpha^2$ versus energy was used for the direct band
gap by extrapolating the linear portions of the curves to zero absorption.
\begin{table}[tbh!]
\caption{Band gap values of $100$-nm-thick pristine TiO$_2$ and SnO$_2$
films and of TiO$_2$ and SnO$_2$ films irradiated by a 200-MeV Ag$^{+15}$ ion beam.}\label{tab1}
\begin{center}
\begin{tabular}{c|c|c}
\hline \hline
\multirow{1}{*}{Irradiation fluence} & \multicolumn{2}{c}{Band gap values}
\\ \cline{2-3}
\multirow{1}{*}{(ions/cm$^2$) } & TiO$_2$ & SnO$_2$ \\ \hline
Pristine & 3.44 & 3.92 \\
$1\times10^{11}$ & 3.57 & 3.78 \\
$1\times10^{12}$ & 3.50 & 3.86 \\
$5\times10^{12}$ & 3.59 & 3.60 \\ \hline\hline
\end{tabular}\end{center}\end{table}

\begin{figure}[tbh!] \centering
\includegraphics[width=6.5cm]{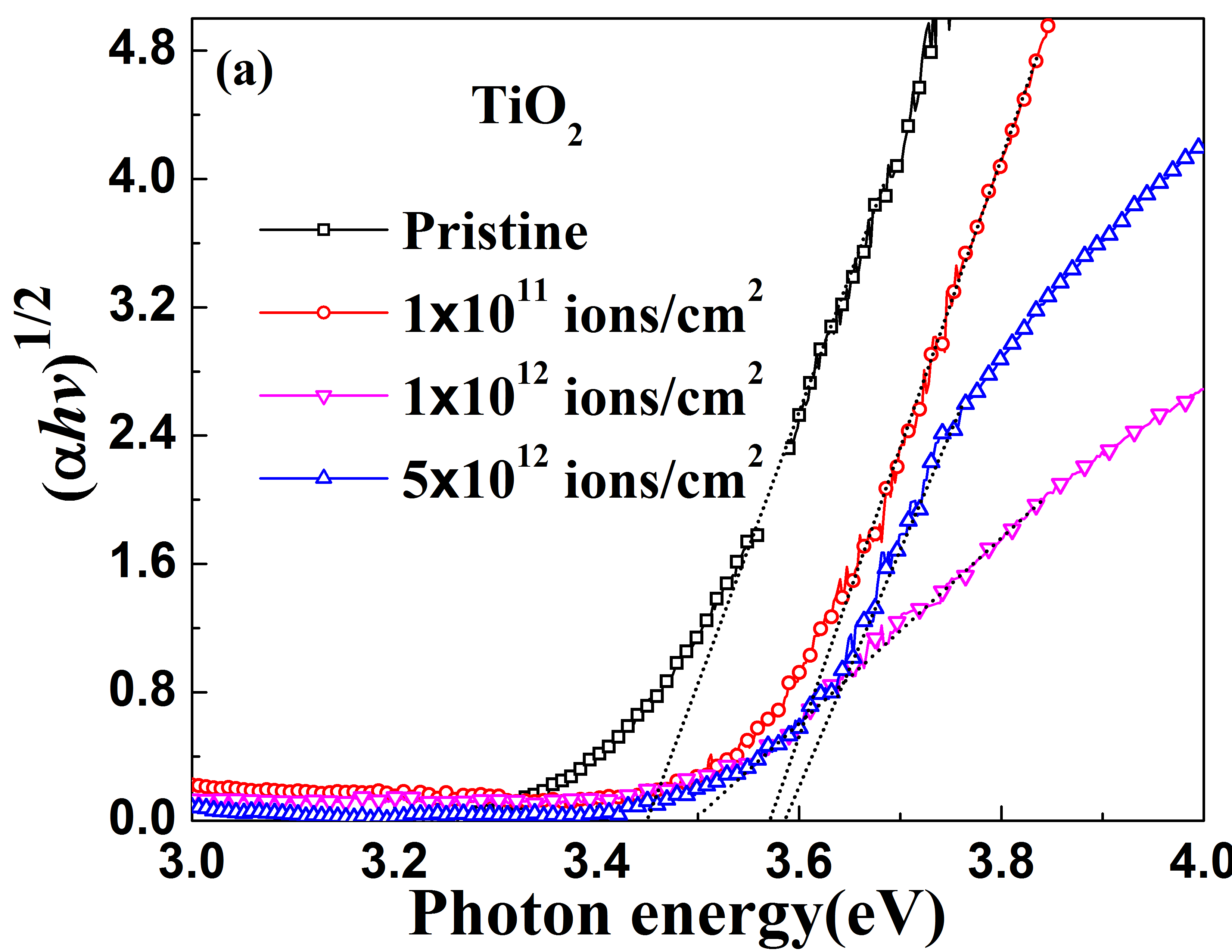}\\  %
\includegraphics[width=6.5cm]{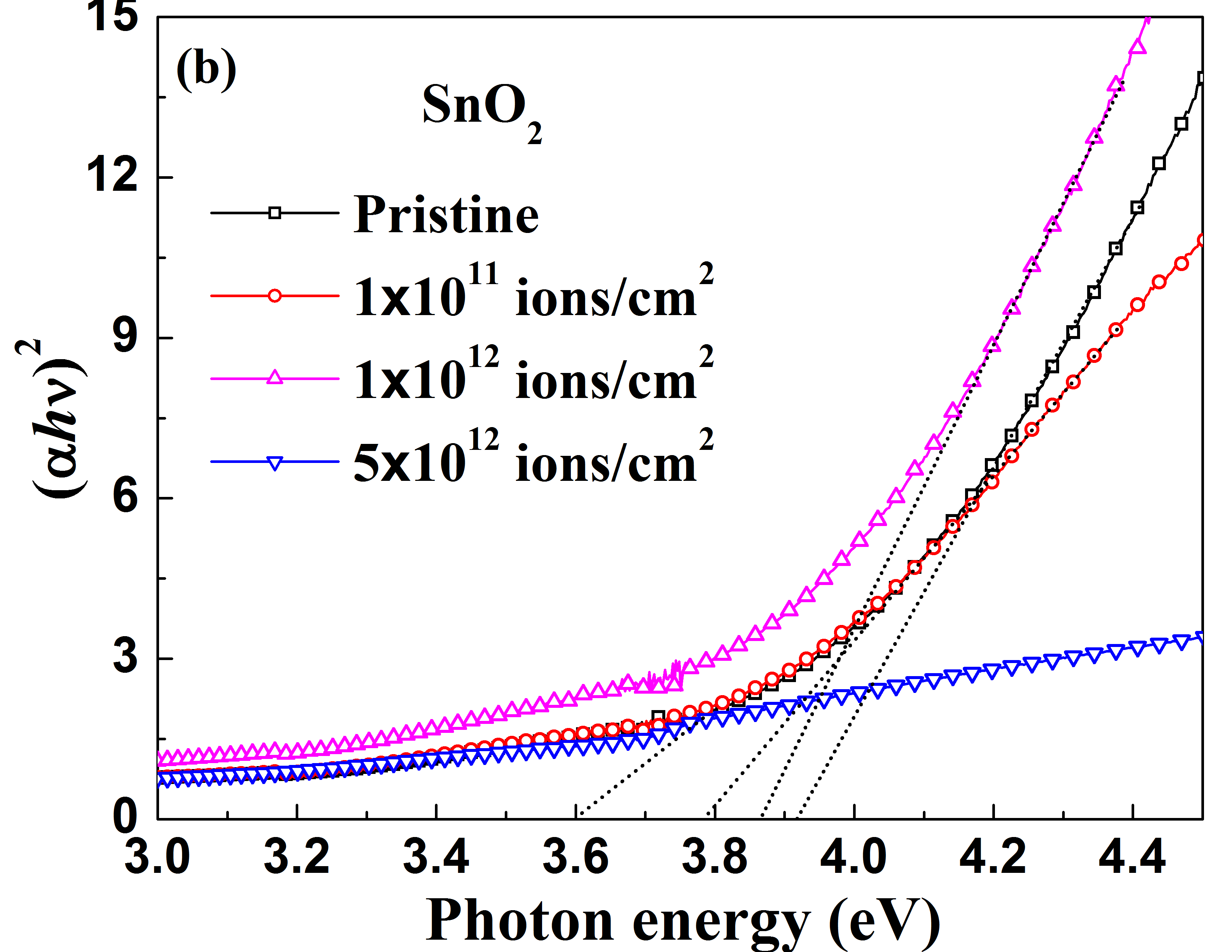}\\
\caption{(Color online) Plots of (a) $(\protect\alpha h \protect\nu)^{1/2}$ versus
$h\protect\nu$ for the pristine and the SHI-irradiated TiO$_2$ thin films and (b)
$(\protect\alpha h\protect\nu)^2$ versus $h\protect\nu$ for the pristine and the
SHI-irradiated SnO$_2$ thin films collected at RT.}\label{fig4}
\end{figure}

The variations in the band gaps for the pristine and the SHI-irradiated TiO$_2$
thin films are depicted in the Fig. \ref{fig4}(a). The optical transitions for TiO$_2$
have been shown to be predominantly indirect \cite{r20,r21} while that for
SnO$_2$ is direct \cite{r22}. The overall values obtained for all the
irradiation fluences are higher than the reported values for TiO$_2$, which are
usually reported to be around $3.2$ eV. The highest band gap value ($\sim 3.6$ eV) was obtained
for the sample irradiated at a fluence of $5\times 10^{12}$ ions/cm$^2$.  The
band gap values increased until a SHI fluence of $1\times10^{12}$ ions/cm$^2$%
, after which they  decreased slightly from the value for the  sample irradiated at a SHI fluence
of $1\times10^{11}$ ions/cm$^2$, $3.6-3.5$ eV. For the irradiated SnO$_2$
samples (see Fig. \ref{fig4}(b)), a  trend similar to that  of the irradiated TiO$_2$ samples is followed.
The highest value for the band gap of the SnO$_2$ sample is for the pristine sample with a value of $3.92$ eV.
The values decreased until a SHI fluence of $1\times10^{12}$ ions/cm$^2$, after which the value
increased slightly to $3.86$ eV. For the sample irradiated at a fluence of $5\times10^{12}$ ions/cm$^2$,
the value again decreases to $3.6$ eV. Table \ref{tab1} shows the variations in the band gap energy with irradiation
fluence for both the TiO$_2$ and the SnO$_2$ systems. HRXRD data indicate that SHIs create controlled structural disorder in the lattices of oxide materials. This can be responsible for the generation of defect levels near the
conduction band, i.e., shallow energy levels, which can give rise to a
transition from the valence band to those levels instead of  band-to-band
transitions. Due to the shallow levels, the band gap is effectively changed. This
decrease in the band gap gives an indication of the stoichometric deviation
of the irradiated SnO$_2$ and the increase in the oxygen vacancies in the SnO$_2$
lattice. A similar behavior was also reported in a previous work \cite{r23}%
, where the effect of In-doping concentration on the optical band gap of
nano-SnO$_2$ was investigated as a function of the calcination temperature.
Another possible explanation for the changes in the band gap value may be
due to the fact that the density of surface states induced in the SnO$_2$ lattice
is modified on irradiation.
\begin{figure}[tbh!]\centering
\includegraphics[width=6.5cm]{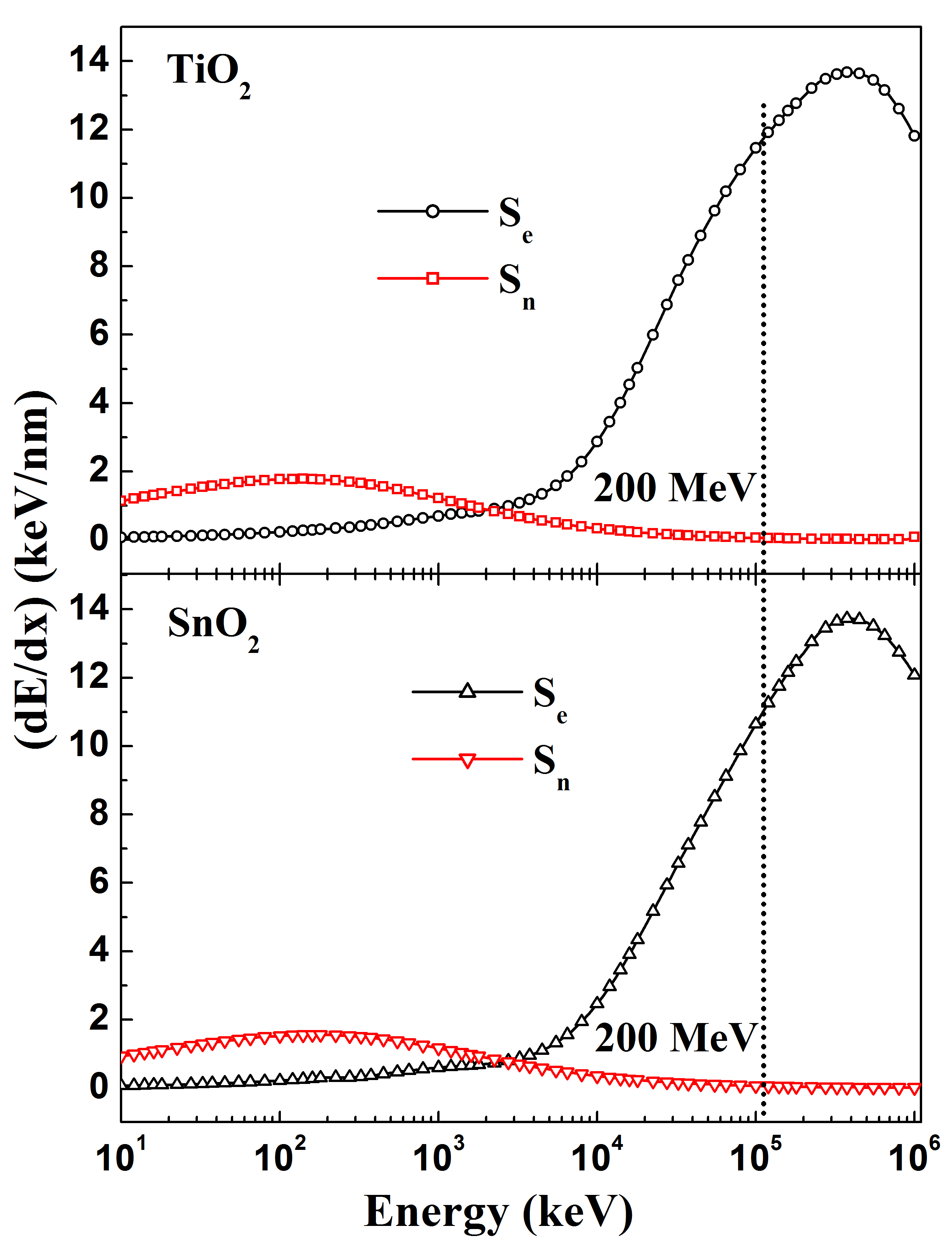}\\
\caption{(Color online) Electronic and nuclear energy losses of 200-MeV Ag$^{+15}$ ions
as functions of the ion energy inside TiO$_2$ and SnO$_2$ targets.}
\label{fig5}\end{figure}

The values of optical absorption edges obtained in our
experiments are observed to deviate from the known values reported in the literature \cite{r9,r24}.
The obvious band gap energy changes of  both oxide systems indicate the possibility of band gap engineering in the as-deposited thin films by means of SHI irradiation. If the observed modifications
in the optical properties are to be understood, an analysis of the possible
implications of ion transport through the film is necessary. When 200-MeV Ag$^{15+}$ ions
pass through the oxide films, they loses energy by collisions with nuclei
(nuclear stopping power S$_n$) and inelastic collisions with electrons
(electronic stopping power S$_e$). In the present case, for a TiO$_2$ film,  S$_e\sim 12.96$ keV/nm and S$_n \sim 32.30$ eV/nm while for a SnO$_2$ film,  S$_e$ $\sim 12.72$ keV/nm and S$_n \sim 36$ eV/nm as calculated
by using stopping and range of ions in matter software \cite{r25}. The vertical
line marks a 200-MeV  incident ion energy, the energy used in the
present work (see Fig. \ref{fig5}). From these values, at higher energies, the electronic energy loss can be seen to dominate the nuclear energy loss, and at an energy of a few hundreds of keV, the opposite is true. Thus, the inelastic electronic collision process is the dominant energy loss mechanism, which induces point and columnar defects and can lead to an increase in the defect density and to a modification in the lattice structure \cite{r26}.

\section{Conclusions}

In this study, the transmittance of the TiO$_2$ thin films increased with increasing SHI fluence while the transmittance of the SnO$_2$ thin films decreased. Both oxide systems showed improvements in the transmittance as the irradiation fluence was increased. An anomalous trend was observed for the optical absorption edges with increasing SHI fluence for both the oxide materials. For irradiated TiO$_2$ thin films,  the optical band gap values for the indirect transitions were shown to be much higher than the expected values reported. The highest value of the band gap was achieved in the sample SHI-irradiated at a fluence of $5\times10^{12}$ ions/cm$^2$ while the lowest value of the band gap was observed in the pristine sample. A similar trend was observed  in the irradiated SnO$_2$ samples. However, for the irradiated SnO$_2$ thin films, the band gap values on an average decreased.

From the results, we can conclude that SHI does not critically affects the pristine TiO$_2$ samples while for SnO$_2$, significant modifications in the optical properties are observed. These observed changes in the optical properties of both the pristine TiO$_2$ and SnO$_2$ films with SHI fluence can be attributed to controlled structural disorder/defects in the system. Our results show a direct linkage between SHI-induced structural disorder/defects and modifications in the optical properties of the oxide materials.

\begin{acknowledgments}
The authors would like to thank the Inter-University Accelerator Centre, New Delhi, India and the Korea Institute of Science and Technology (KIST, Project No. 2V02631), Seoul, Korea for experimental support. The Department of Science and Technology (DST), Government of India, is acknowledged for supporting this work under Project No. S2/SR/CMP-0051/2007.
\end{acknowledgments}

\end{document}